\documentclass[%
 reprint,
 amsmath,amssymb,
 aps,
]{revtex4-2}

\usepackage{graphicx}
\usepackage{dcolumn}
\usepackage{bm}
\usepackage{textcomp}
\usepackage{natbib} 

\usepackage{textcomp}
\usepackage[outdir=./]{epstopdf}
\usepackage{tabularx} 
\usepackage{times}
\usepackage{wrapfig}
\usepackage{xcolor}
\usepackage{caption}
\usepackage{subcaption}
\usepackage{adjustbox}
\usepackage[most]{tcolorbox}
\usepackage{listings}
\usepackage{silence}
\usepackage{wasysym}
\usepackage{titlesec}
\begin{document}

\preprint{APS/123-QED}

\title{Ab initio simulation of amorphous graphite.}

\author{R. Thapa}
\email{rt887917@ohio.edu}
\affiliation{Department of Physics and Astronomy, \\
Nanoscale and Quantum Phenomena Institute (NQPI),\\
Ohio University, Athens, Ohio 45701, USA}%

\author{C. Ugwumadu}
\email{cu884120@ohio.edu}
\affiliation{Department of Physics and Astronomy, \\
Nanoscale and Quantum Phenomena Institute (NQPI),\\
Ohio University, Athens, Ohio 45701, USA}%

\author{K. Nepal}
\email{kn478619@ohio.edu}
\affiliation{Department of Physics and Astronomy, \\
Nanoscale and Quantum Phenomena Institute (NQPI),\\
Ohio University, Athens, Ohio 45701, USA}%

\author{J. Trembly}%
\email{trembly@ohio.edu}
\affiliation{Department of Mechanical Engineering\\
                 Institute of Sustainable Energy and the Environment, \\
Ohio University, Athens, Ohio 45701, USA}%

\author{D. A. Drabold}%
\email{drabold@ohio.edu}
\affiliation{Department of Physics and Astronomy, \\
Ohio University, Athens, Ohio 45701, USA}%

\date{\today}

\begin{abstract}
\noindent An amorphous graphite (a-G) material has been predicted from molecular dynamics simulation using {\it ab initio} methods. Carbon materials reveal a strong proclivity to convert into a sp$^2$ network and then layer at temperatures near 3000K within a density range of ca. 2.2-2.8 g/cm$^3$. Each layer of a-G is a monolayer of amorphous graphene including pentagons and heptagons in addition to hexagons, and the planes are separated by about 3.1 $\mathring{A}$. The layering transition has been studied using various structural and dynamical analyses. The transition is unique as one of partial ordering (long range order of planes and galleries, but topological disorder in the planes). The planes are quite flat, even though monolayer amorphous graphene puckers near pentagonal sites. Inter-plane cohesion is due partly to non Van der Waals interactions. The structural disorder has been studied closely, especially the consequences of disorder to electronic transport. It is expected that the transition elucidated here may be salient to other layered materials.
\end{abstract}

\maketitle

Carbon-based materials seem to have unlimited potential applications and interest \cite{Cha13, Bafkary16, Meyyappan16}, from life to Bucky Balls, and they continue to yield scientific surprises and new applications. 

Graphite is an important, commonly available carbon material with many uses. A burgeoning application for graphite is for battery electrodes in Li ion batteries \cite{Asenbauer2020} and is crucial for the EV industry -- a Tesla model S on average needs 54 kg of graphite \cite{Tesla}. Such electrodes are best if made with pure carbon materials, which are becoming more difficult to obtain owing to spiraling technological demand. It is therefore of interest to determine novel paths to synthetic forms of graphite from naturally occurring carbonaceous material such as coal. This raises several questions: (1) Is it possible to convert such materials into a graphitic phase?; (2) What impurities will remain and with what technological consequences?; (3) What are the resulting properties (structural, mechanical, electrical and thermal) of such materials? 

In a series of papers, we have discussed an amorphous phase of monolayer graphene, based on structural models involving pure sp$^2$ bonding with ring disorder (that is, rather than a 2D net consisting only of hexagons, we allow for pentagons, heptagons, etc.). Among other findings, we noted that the presence of pentagons in such a structure induces puckering (departure from ideal planarity) from the strain of the ring defect using {\it ab initio} methods \cite{Li11, Li12}. The semi-metallic character of perfect graphene is transformed by ring disorder \cite{Kapko10, Van12}. Recently, experimental synthesis of monolayer amorphous graphene using chemical vapor deposition has been reported \cite{Toh20}. On the theoretical side, two dimensional amorphous graphene structures created by quenching the high temperature liquid state using Tersoff-II \cite{PRB1988} potential has been reported \cite{Avishek12}. Graphitization of amorphous carbon under  electron irradiation has been studied experimentally and theoretically \cite{Felix2012}.

In this Letter, we employ an {\it ab initio} method to unveil a layering transition from either amorphous carbon or even {\it random} starting models into a structure consisting of planes of monolayer amorphous graphene separated by $\sim$0.3 nm, the interlayer separation in graphite, as a consequence of annealing such models with first principles interactions at a temperature of about 3000K, and for a density range of ca. 2.2-2.8 g/cm$^3$). These  sheets are sp$^2$, but with ring disorder (pentagons, hexagons, heptagons). We name this  material Amorphous Graphite (a-G). We elucidate the transition in atomistic detail. The a-G structure cannot exactly reproduce AB stacking, yet even with ring disorder on the planes, has a total energy only ~0.32 eV/atom above crystalline graphite (c-G). We examine the electronic origins of cohesion by analyzing electronic structure in the galleries, and demonstrate how ring disorder reduces in-plane electron transport. This paper takes a step toward realizing the goal of synthetic graphite, and may offer clues to layering processes in other systems such as metal dichalcogenides. It has been suspected from experiments that graphitization occurs near 3000K  \cite{SCHUPFER2021,Barreiro13, Westenfelder11, Kim16, Acheson1931}, but the details of the formation process and nature of disorder in the planes was unknown. From a modeling viewpoint, {\it ab initio} simulations of complete or partial ordering are rare and important, the best example being phase-change memory materials \cite{Hegedus08, Kalikka16}.

\begin{figure}
\begin{adjustbox}{varwidth = 0.48\textwidth, center,   rndfbox= 3ex 3ex 3ex 3ex}
	\begin{subfigure}{.95\textwidth}
  		\centering
 		 \tcbox[top = 0.5pt, left = 0.5pt, right = 0.5pt, bottom = 0.5pt]{ \includegraphics[width=.95\linewidth]{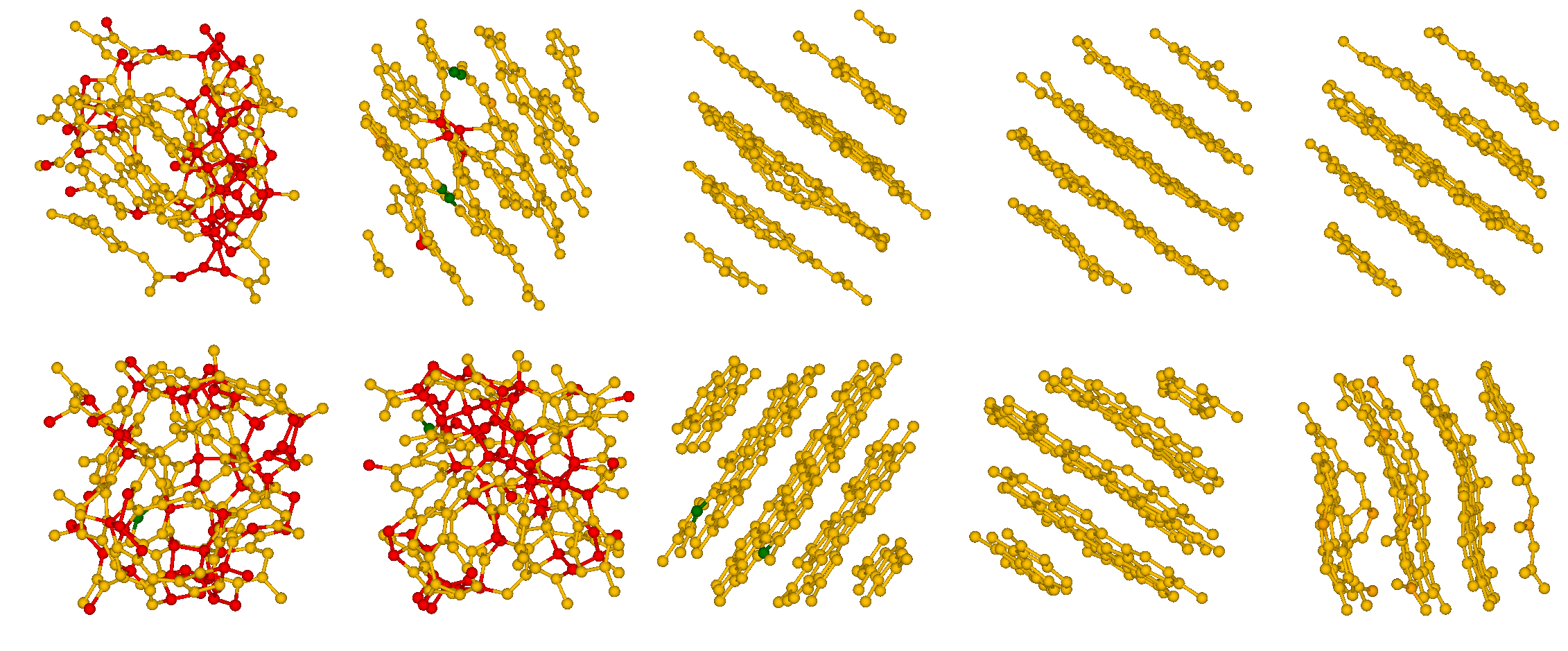}  }
	\end{subfigure}
	\newline
	\begin{subfigure}{.95\textwidth}
		  \centering
 		 \tcbox[top = 2.5pt, left = 2.5pt, right = 2.5pt, bottom = 2.5pt]{ \includegraphics[width=.95\linewidth]{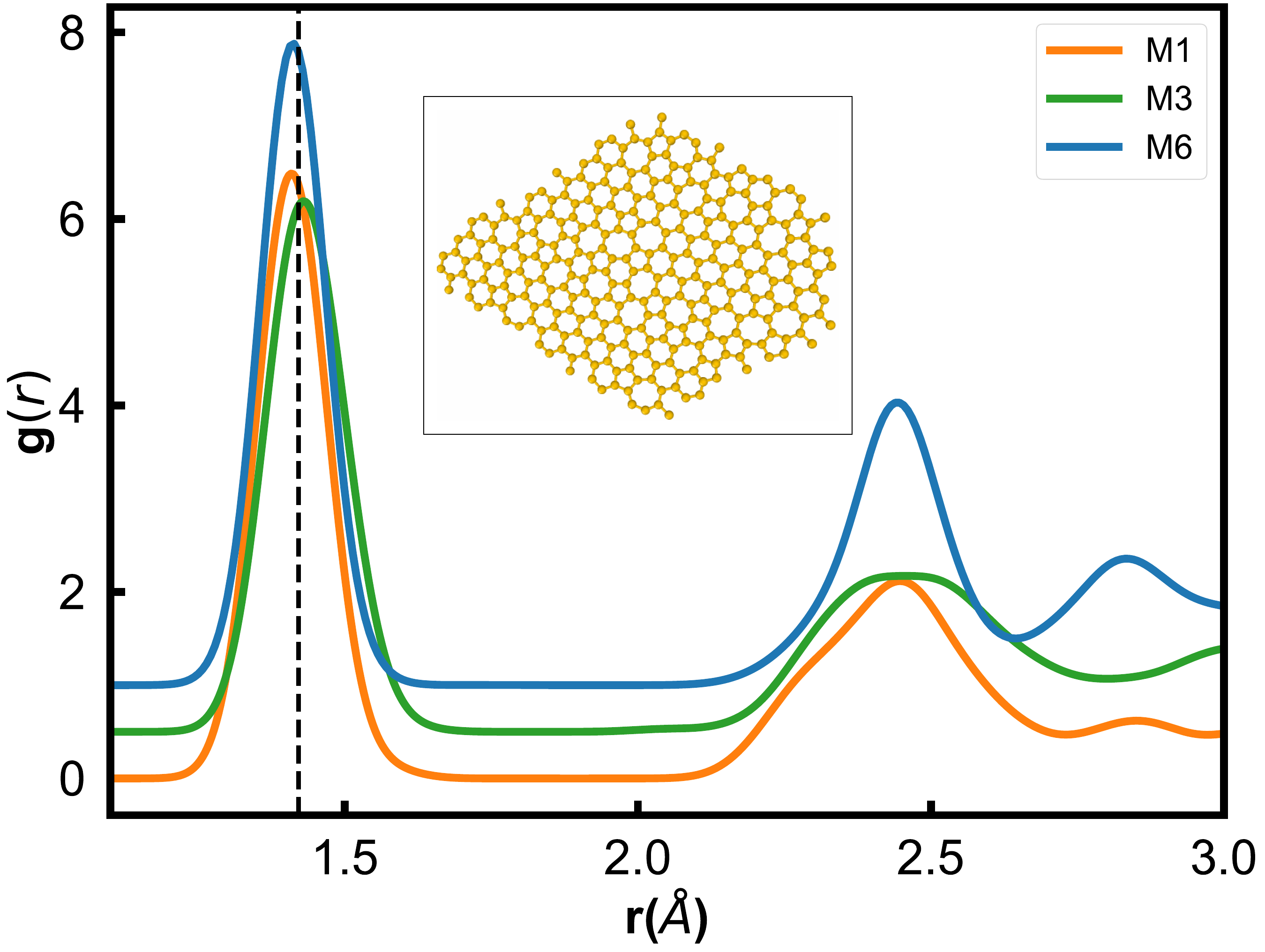}  }
	\end{subfigure}
	\caption{\textbf{(top)} Conjugate gradient relaxed structure of M1 (top) and M2 (bottom) after NVT simulation at 300K, 2500K , 2700K, 3000K, 3300K. \newline \textbf{(bottom)} In-plane RDF of the representative models. The dashed line indicates the graphite 
	bond length. The inset shows the arrangement of atoms in a representative layer in a a-G. }
	\label{fig:CONTCAR}
\end{adjustbox} 
\end{figure}
\begin{figure}
\begin{adjustbox}{varwidth = 0.48\textwidth, center,   rndfbox= 3ex 3ex 3ex 3ex}
	\begin{subfigure}{.8\textwidth}
  		\centering
     		 \tcbox[top = 0.5pt, left = 0.5pt, right = 0.5pt, bottom = 0.5pt]{ \includegraphics[width=.95\linewidth]{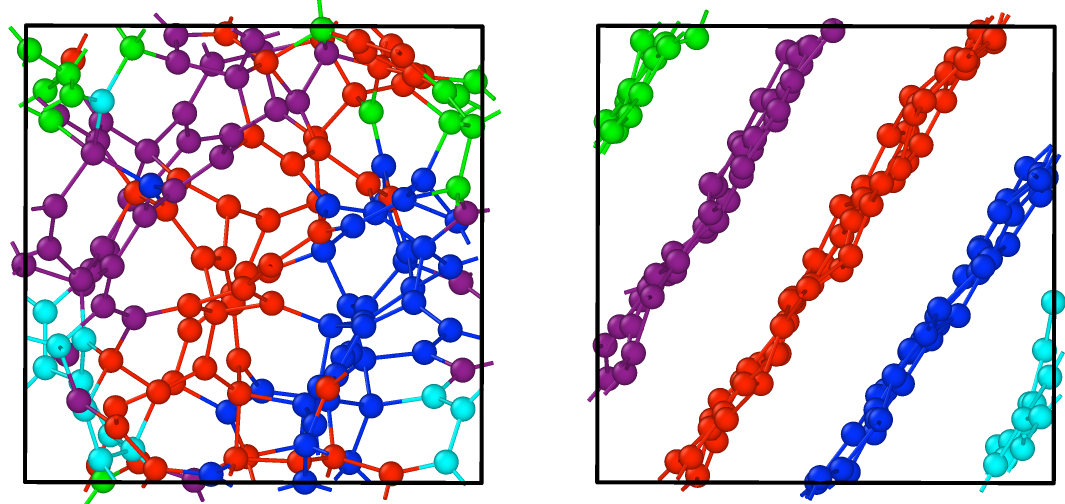}  }
	\end{subfigure}
	\begin{subfigure}{.95\textwidth}
		  \centering
 		 \tcbox[top = 2.5pt, left = 2.5pt, right = 2.5pt, bottom = 2.5pt]{ \includegraphics[width=.95\linewidth]{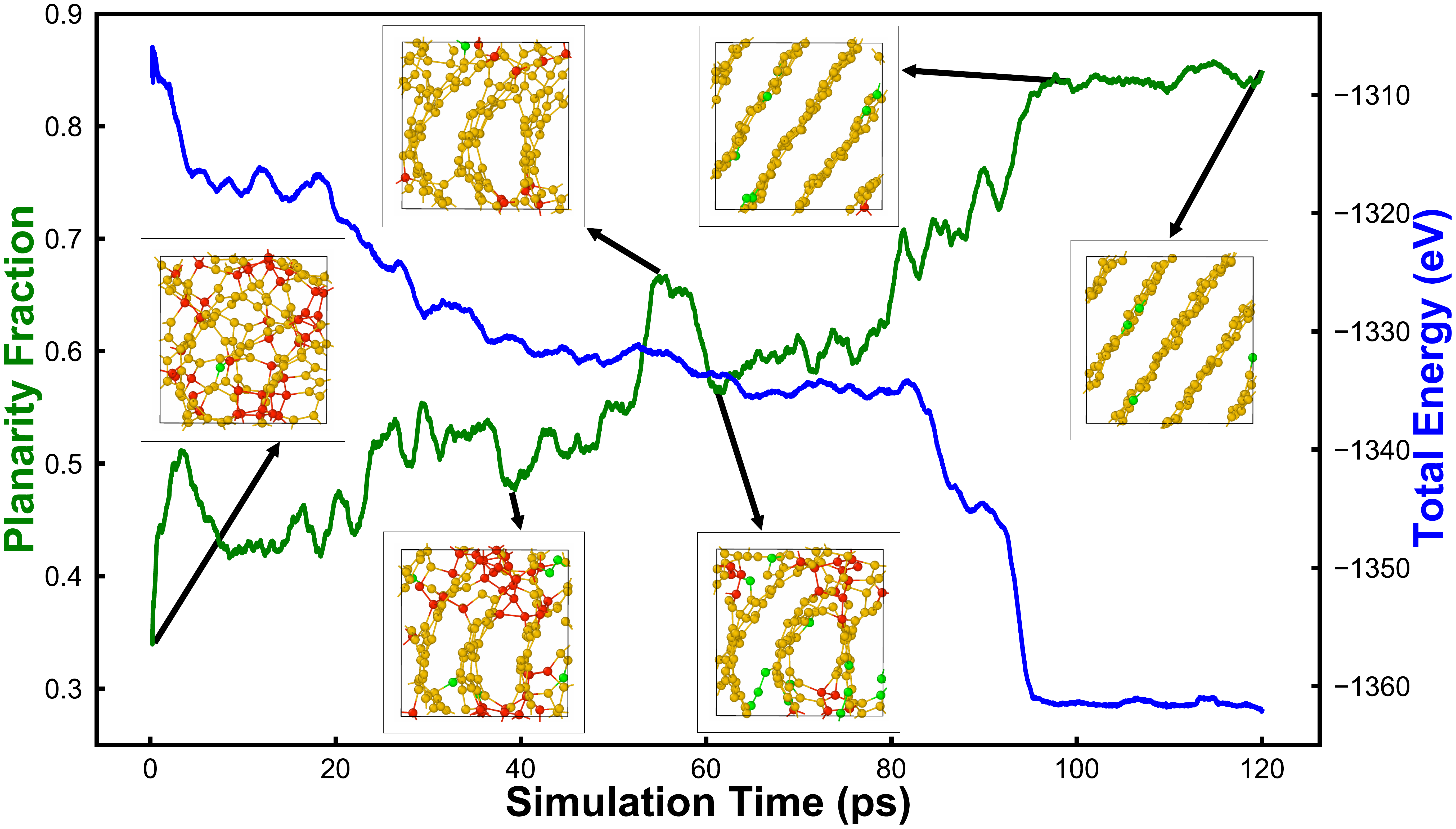}  }
	\end{subfigure}
	\caption{\centering \textbf{(top)} Positions of atoms forming different layers in a-G for M2 at 2700K. Atoms forming different layers are shown in different colors. \newline 	\textbf{(bottom)}  Layer       formation and total energy (plotted as a moving average over 2 ps) as a function of simulation time. The insets show the snapshots of the 	atomic configurations at different points in time.}
	\label{fig:Layers}
\end{adjustbox}
\end{figure}
\begin{table}
	\begin{ruledtabular}
		\begin{tabular}{cccccr}
			\textrm{Model}&
			\textrm{Size}&
			\textrm{$\tau$ (in ps)}&
			\textrm{$\delta E_{atom}$ (in eV)}&
			\textrm{Functional}&
			\textrm{Initial State}\\
			\colrule
			\rule{0pt}{4mm}
			M1   & 160   & 45   &  0.00  &  PBE & amorphous\\          
			M2	& 160   & 95   &  0.10 & PBE & amorphous\\   
			M3   & 160 & 130  &  -0.04 & PBE+vdW & amorphous \\        
			M4   & 80     & 40   &  0.13  &  PBE & random\\          
			M5	& 400   & 50   &  0.08  & GAP-ML & random\\   
			M6   & 1000  & 60   & -0.04 &  GAP-ML & random\\   
		\end{tabular}
		\caption{Simulation parameters for various models of a-G obtained by NVT simulation at 2700K. Difference in energy per atom, computed with PBE, are compared to M1.}
	    \label{tab:CN}
	\end{ruledtabular}
\end{table}


Our simulation protocol was simple: constant volume simulations were carried out with either (1) {\it ab initio} models of a-C simulated for the selected density \cite{Bishal18} or (2) {\it random} starting configuration for the desired density. These systems were then annealed to T = 3000K using a Nos\'e-Hoover thermostat \cite{Nose1984,Hoover1985}. For a density range of ca. 2.2-2.8 g/cm$^3$ a layering transition was always observed, and the structure maintained the layered structure in subsequent MD simulation, and was topologically unchanged by a conjugate gradient relaxation. Total simulation time ranged from 100 ps to 500 ps. We carried the simulations out with VASP \cite{VASP1996} using projector augmented wave (PAW) \cite{PAW1994} potentials and the {\it Perdew-Burke-Ernzerhof} (PBE) \cite{PBE1996}  exchange-correlation functional. For completeness, we also used the DFT-D3 Van der Waals corrected functional \cite{Grime2010}, and the accurate LDA-trained machine learning GAP potential of Deringer and coworkers \cite{Volker17, Volker21}. These three approaches gave essentially identical results, consistent a-G formation in the density/temperature window. Low density a-C ($<$ 2.0 g/cm$^3$ ) had a significant sp$^3$ to sp$^2$ conversion but weak layering (undulating worm-like layers) while high density a-C ($>$ 3.0 g/cm$^3$ ) did not layer. No layering was seen under simulation at temperatures higher than 4000K. The linear scaling GAP potential enabled much larger simulations than VASP. In contrast, identical simulations with REAX-FF \cite{ReaxFF} or Tersoff \cite{Tersoff88} potentials failed to display layering. In Table I, we summarize the simulations underlying this work. All the calculations listed in the Table employed a density of 2.44 g/cm$^3$. $\tau$ is the simulation time required for layering to become clear. $\tau$ is reasonably consistent over all the simulations and methods for systems including 160-1000 atoms. Finite-size effects were investigated by forming models ranging from 80-1000 atoms, and the GAP potential revealed that essentially an identical layering occurred with comparable $\tau$. This and the consistent form of the a-G implies that our observations are not very sensitive to size effects. We infer that the layering transition temperature is near 2700K, provided the simulation is run for a considerable time ($\sim$ 100ps) with accurate interatomic interactions. A transition temperature of 3000K has been observed experimentally for production of high quality graphene using flash graphene synthesis \cite{Luong19}.

The structural transition of the a-C network from disordered phase into a a-G under NVT simulation at different temperatures are shown in Fig. \ref{fig:CONTCAR} (top) for models M1 and M2. Atoms in the figures are color coded: yellow for sp$^2$, red for sp$^3$ and green for sp. This color nomenclature will be used throughout unless otherwise stated. 
Since c-G is completely sp$^2$ with flat layers, we consider our models to be graphitized into a-G  if it has a significant fraction of sp$^2$ bonding ($>$ 95\% ) and is layered.  Following this definition, we see from Fig. \ref{fig:CONTCAR} (top) that graphitization only happens at and above 2700K in both models with an interplanar separation in the range 3.05 $\pm$ 0.06  $\mathring{A}$. However, there is a significant increase in the fraction of sp$^2$ atoms even at 2500K. This temperature-induced transition from sp$^3$ to sp$^2$ bonding in nanodiamond  and adamantane has been studied experimentally using Raman spectroscopy \cite{SCHUPFER2021}.  Zero pressure relaxation of the a-G models M1 and M2 with vdW interactions produced a lower energy configuration accompanied by an increase in volume. This volume rise lowers the density to 2.15 g/cm$^3$ and increases the interlayer separation to 3.30 $\pm$ 0.05 $\mathring{A}$, notably close to graphite. In contrast to c-G with regular ordering between adjacent layers (AA, AB stacking), there is no such stacking of the layers in a-G, a consequence of the presence of topological (ring) disorder in the planes. In Fig. \ref{fig:CONTCAR} (bottom) we show the in-plane radial distribution functions for the models. The first peak is centered around the graphitic bond length and the width of the peaks arises from disorder-induced deviations in bond length from the ideal graphite bond length. The largest model M6 with 1000 atoms produces extended ordering beyond the first neighbor with clear peaks at 2.45  $\mathring{A}$ and 2.85  $\mathring{A}$.

To study the origin of layering, we tracked where the atoms forming the layers were located in the originally disordered structure: see Fig. \ref{fig:Layers} (top). The atoms in a particular layer of a-G are members of a connected network in the a-C. The disorder-to-order transition seems describable with 
a nucleation theory picture with seeds of sp$^2$ carbon growing into larger planar structures, enabling layering.  

Fig. \ref{fig:Layers} (bottom) treats the time evolution of the transition in detail for M2, similar results were obtained for the other models. The planarity fraction is computed using the odds ratio for the likelihood of atoms forming planes.  The peculiar peaks around 6 ps and 57 ps suggest sharp rearrangement of atomic positions to achieve planar configurations with a higher fraction of sp$^2$ bonding. These configurations underwent additional substantial rearrangements before yielding a-G. It is worth mentioning that after 95 ps the layering order parameter exhibited reduced fluctuations as the system stabilized at an optimized energy, as seen in the flat tail for the total energy curve (in blue). 

\begin{figure*}[!ht]
	\begin{adjustbox}{varwidth = \textwidth, center, rndfbox= 3ex 3ex 3ex 3ex}
    		\begin{subfigure}[b]{.18\textwidth}
 			 \centering
			  \caption*{\Large\textbf{(a)}}
			  \includegraphics[width=\textwidth]{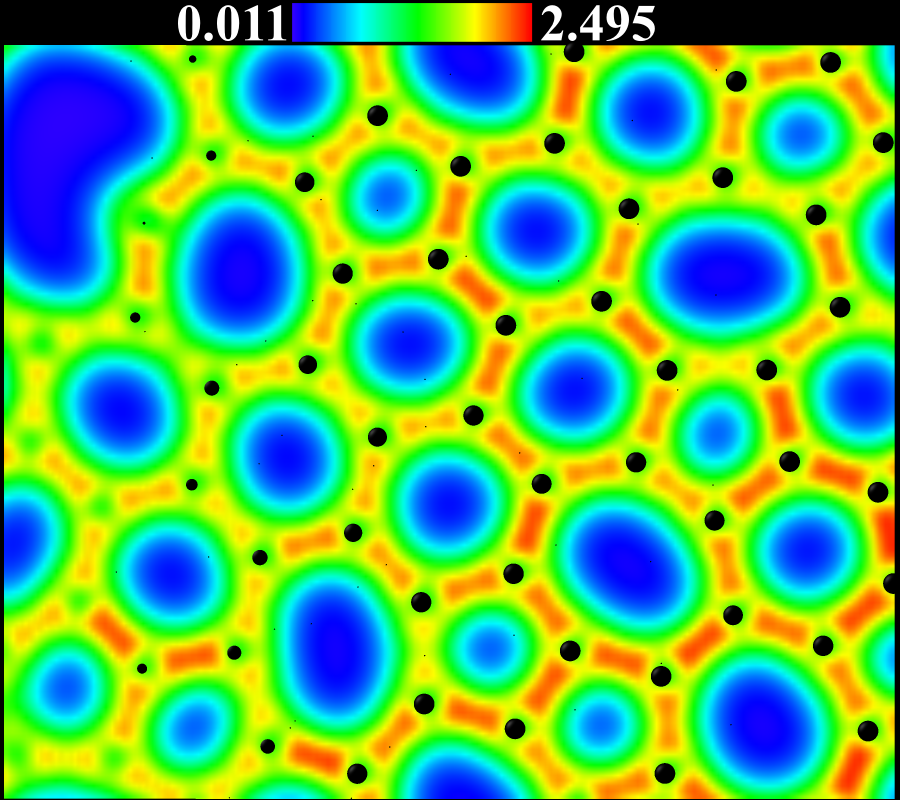}
		\end{subfigure}
		\hspace{0.15cm}
		\begin{subfigure}[b]{.18\textwidth}
 			 \centering
  			\caption*{\Large\textbf{(b)}}
 			 \includegraphics[width=\textwidth]{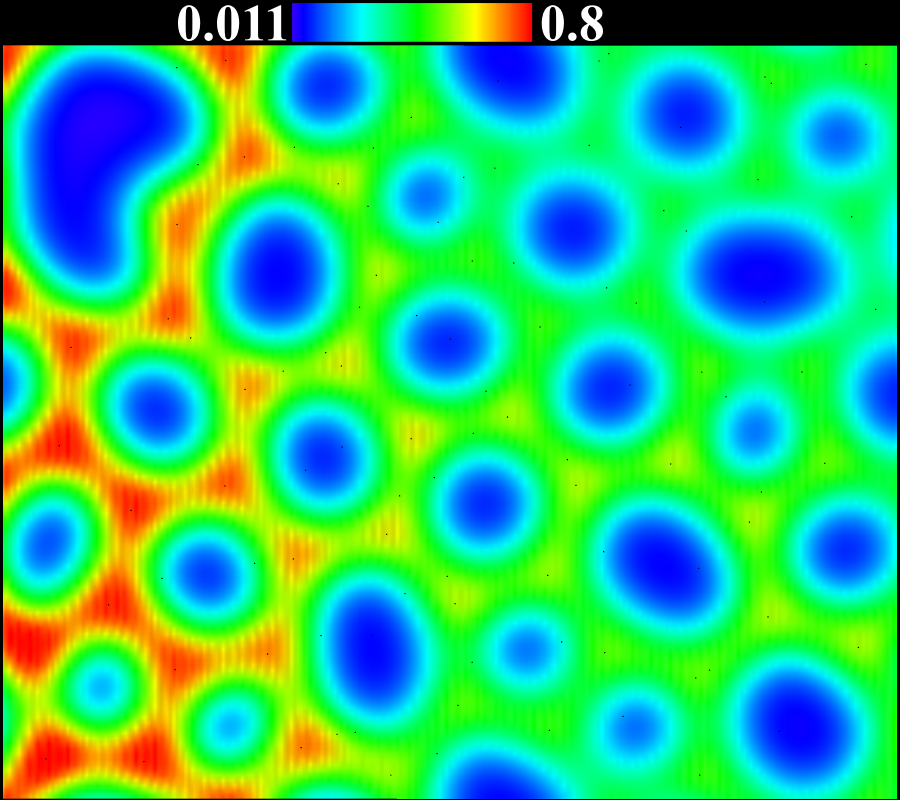}
		\end{subfigure}
		\hspace{0.15cm}
		\begin{subfigure}[b]{.18\textwidth}
 			 \centering
 			 \caption*{\Large\textbf{(c)}}
 			 \includegraphics[width=\textwidth]{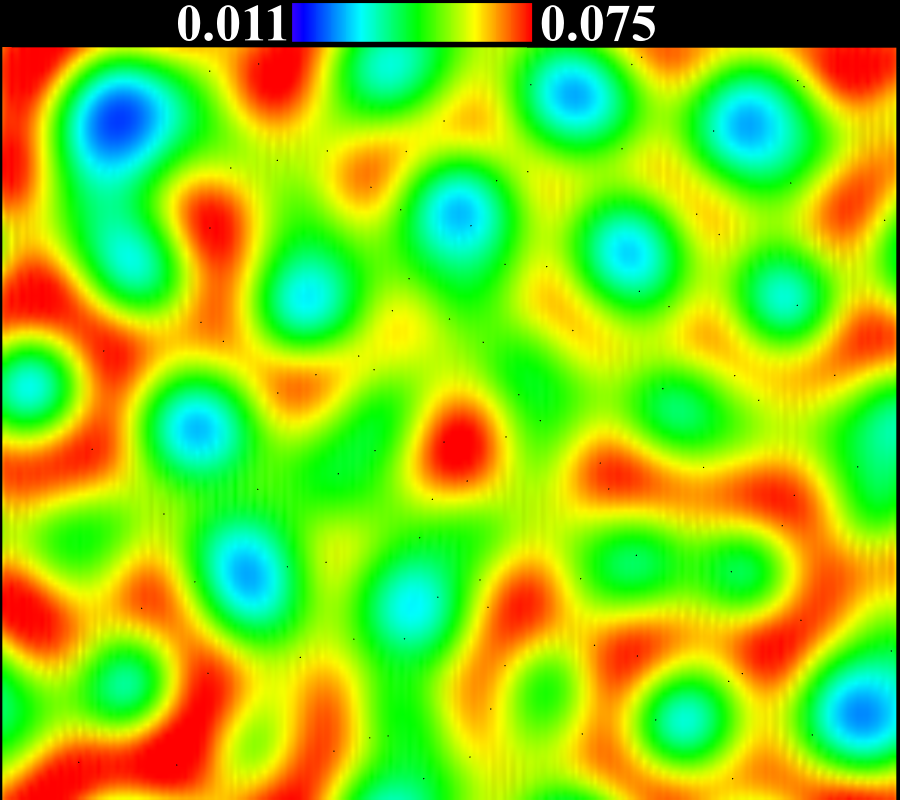}
		\end{subfigure}
		\hspace{0.15cm}
		\begin{subfigure}[b]{.18\textwidth}
			  \centering
 			 \caption*{\Large\textbf{(d)}}
			  \includegraphics[width=\textwidth]{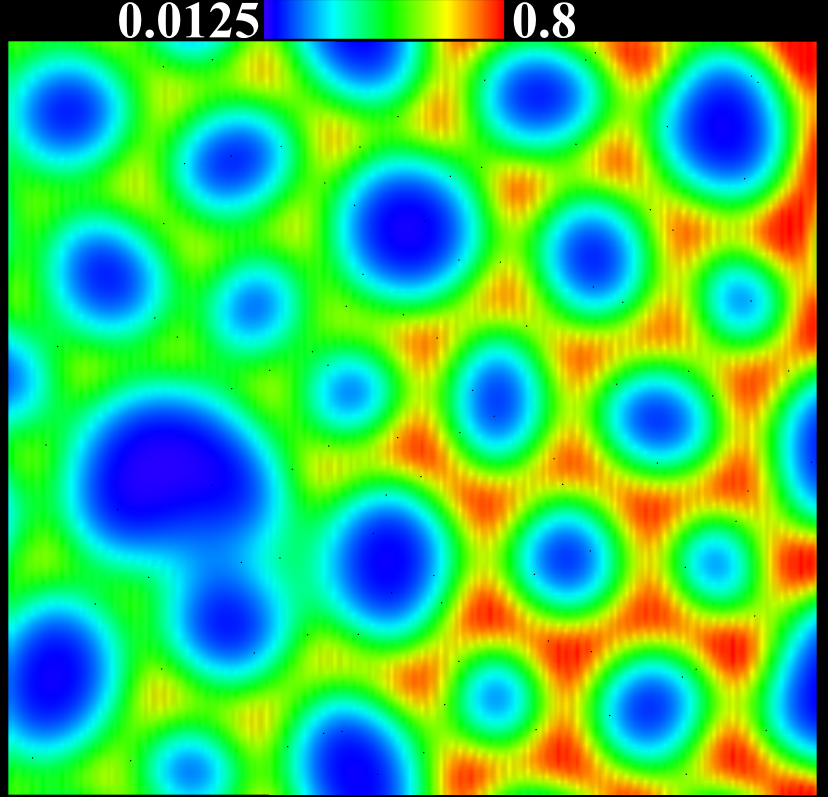}
		\end{subfigure}
		\hspace{0.15cm}
		\begin{subfigure}[b]{.18\textwidth}
 			 \centering
  			\caption*{\Large\textbf{(e)}}
  			\includegraphics[width=\textwidth]{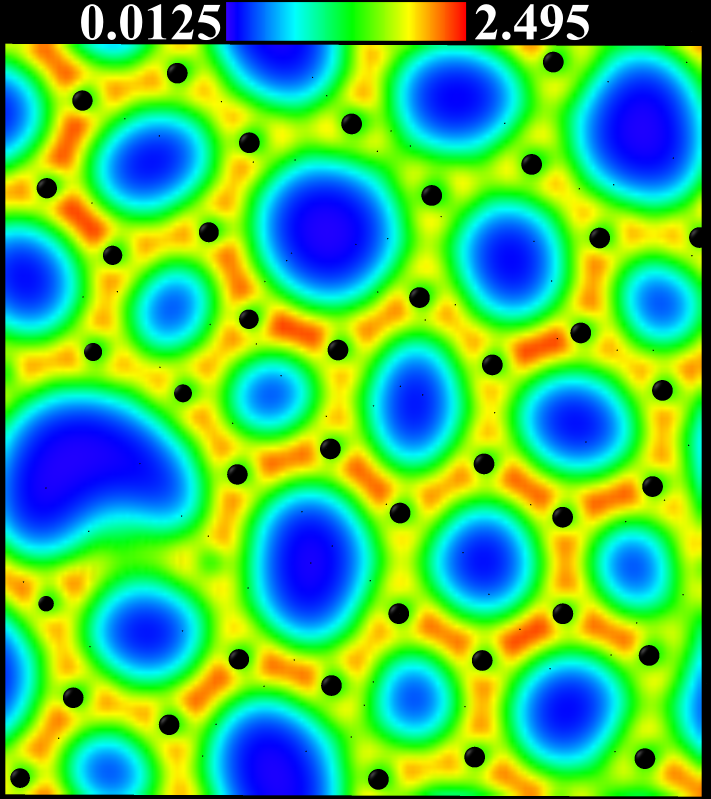}
		\end{subfigure}

		\vspace{0.15cm}

		\begin{subfigure}[b]{.18\textwidth}
			  \centering
  			\includegraphics[width=\textwidth]{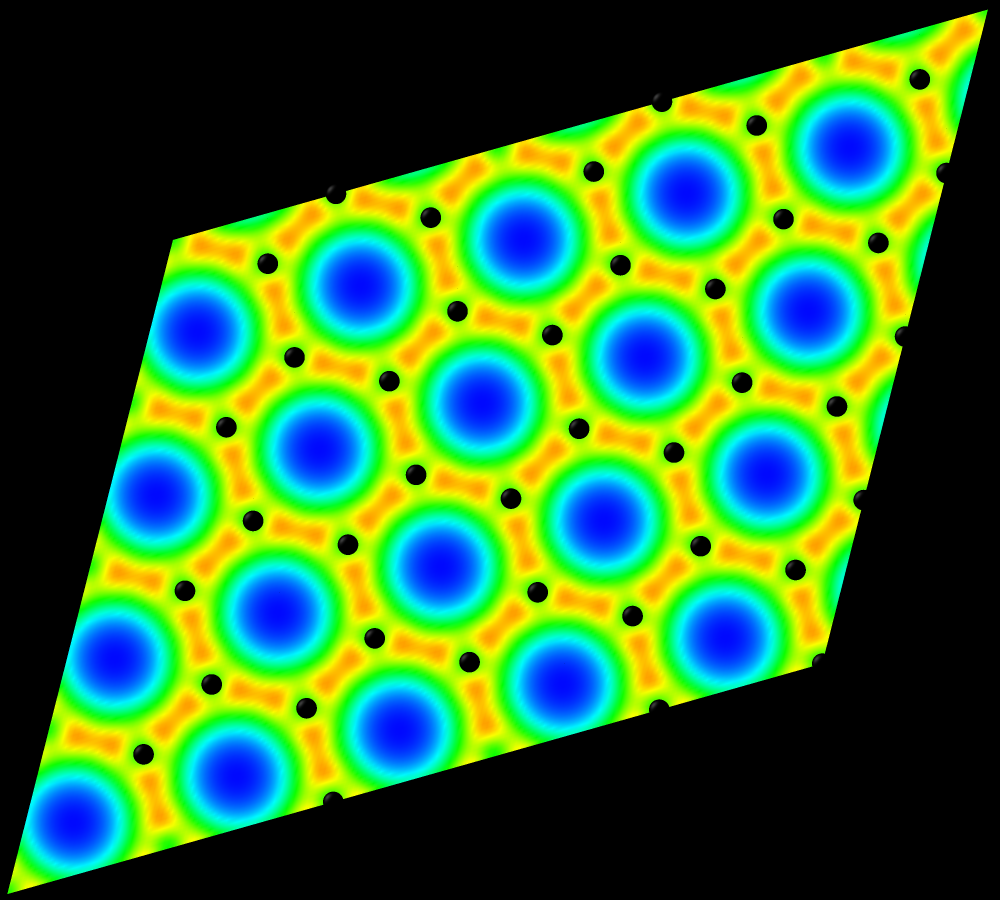}
			\end{subfigure}
		\hspace{0.15cm}
		\begin{subfigure}[b]{.18\textwidth}
 			 \centering
 			 \includegraphics[width=\textwidth]{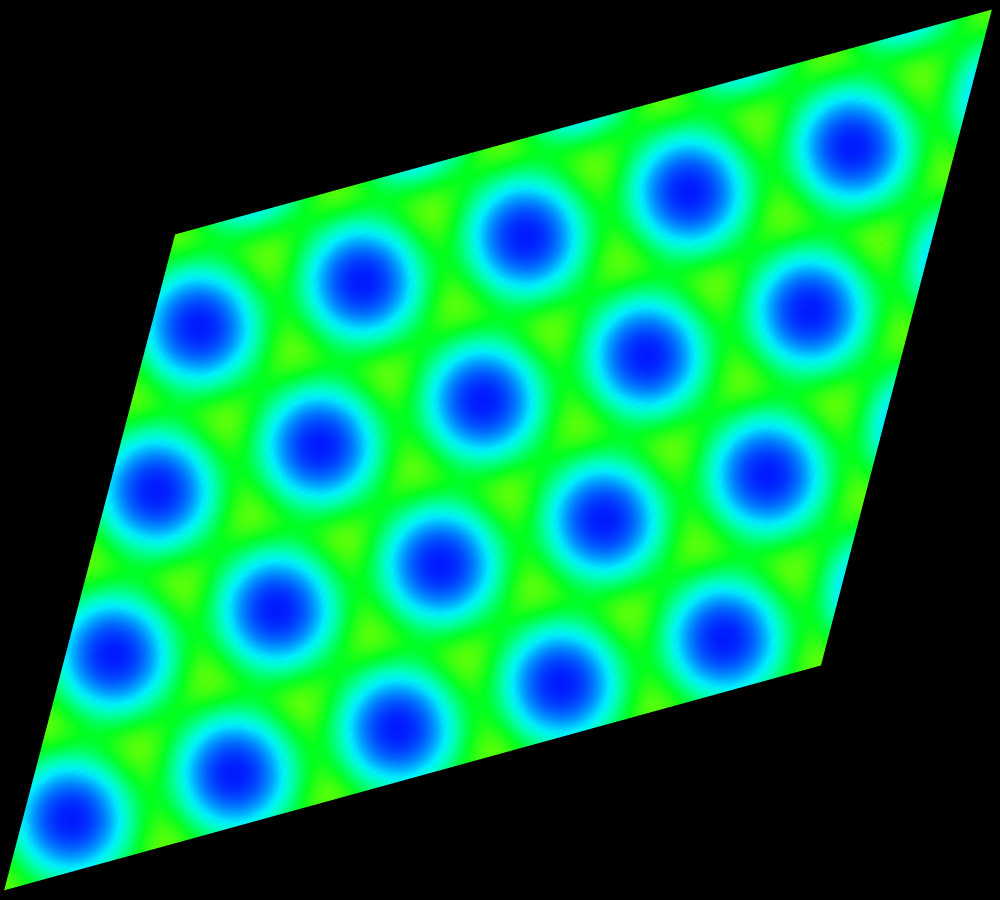}
		\end{subfigure}
		\hspace{0.15cm}
		\begin{subfigure}[b]{.18\textwidth}
			  \centering
 			 \includegraphics[width=\textwidth]{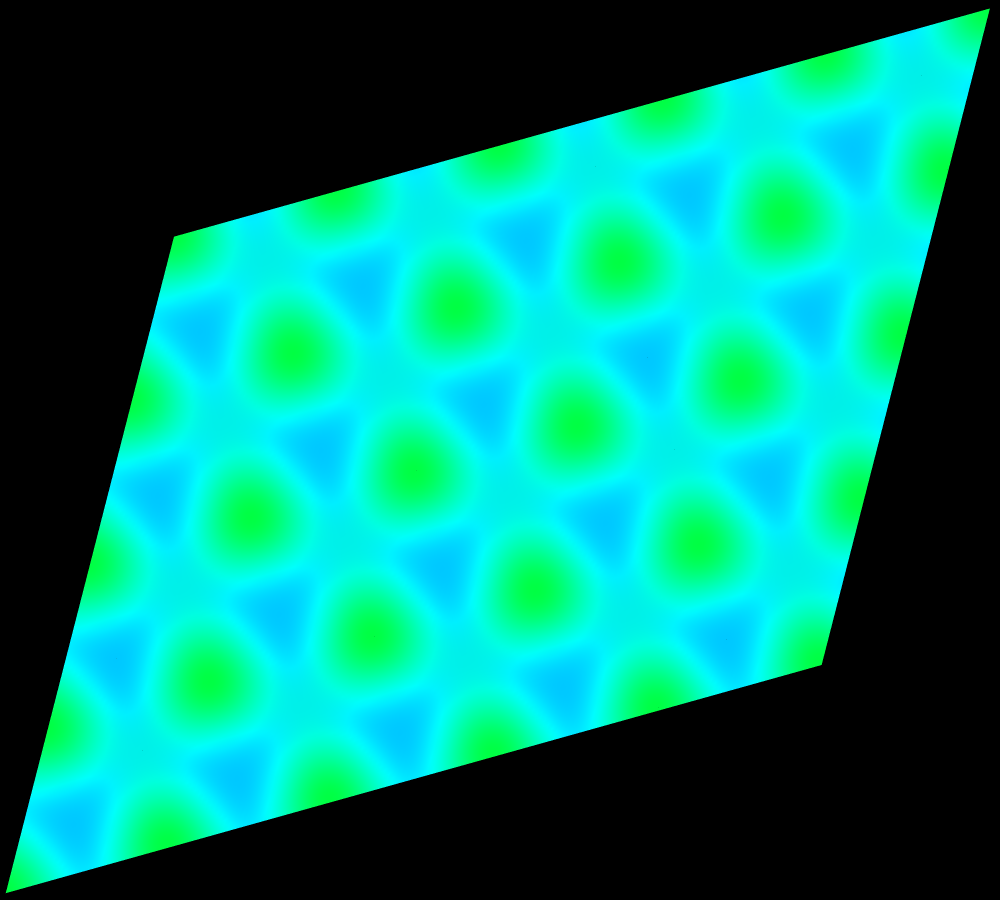}
			\end{subfigure}
		\hspace{0.15cm}
		\begin{subfigure}[b]{.18\textwidth}
 			 \centering
 			 \includegraphics[width=\textwidth]{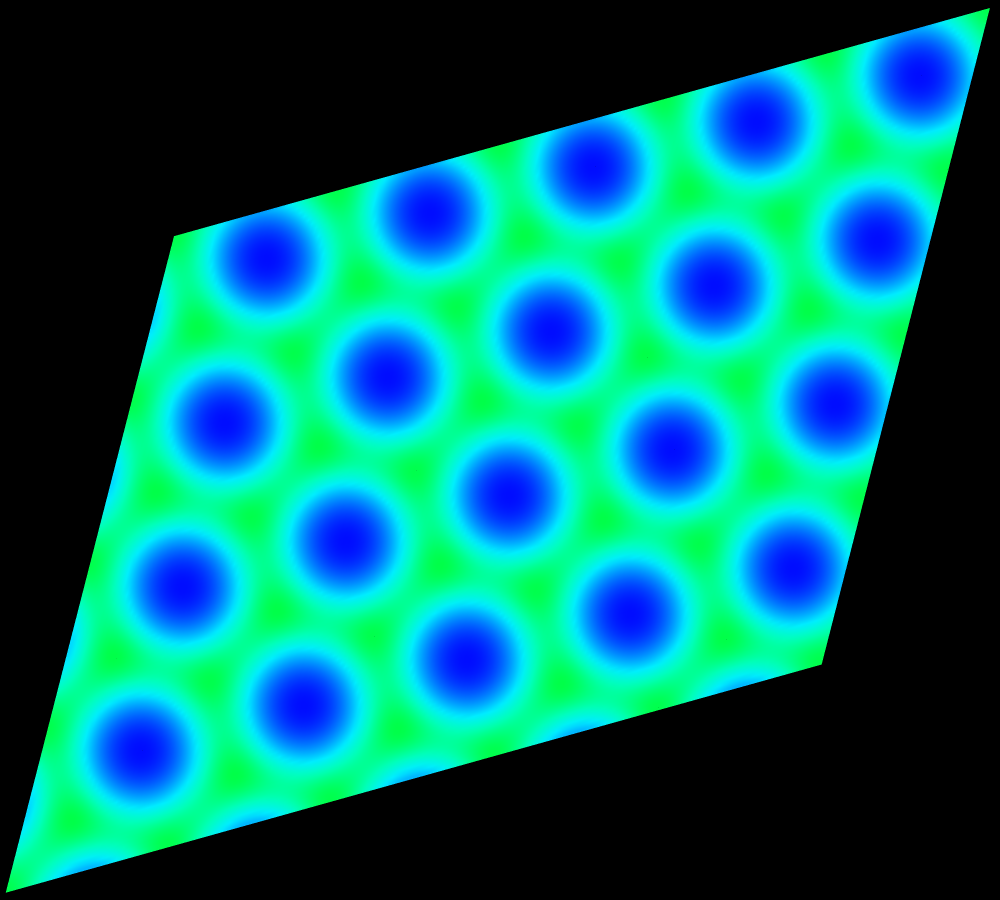}
		\end{subfigure}
		\hspace{0.15cm}
		\begin{subfigure}[b]{.18\textwidth}
 			 \centering
 			 \includegraphics[width=\textwidth]{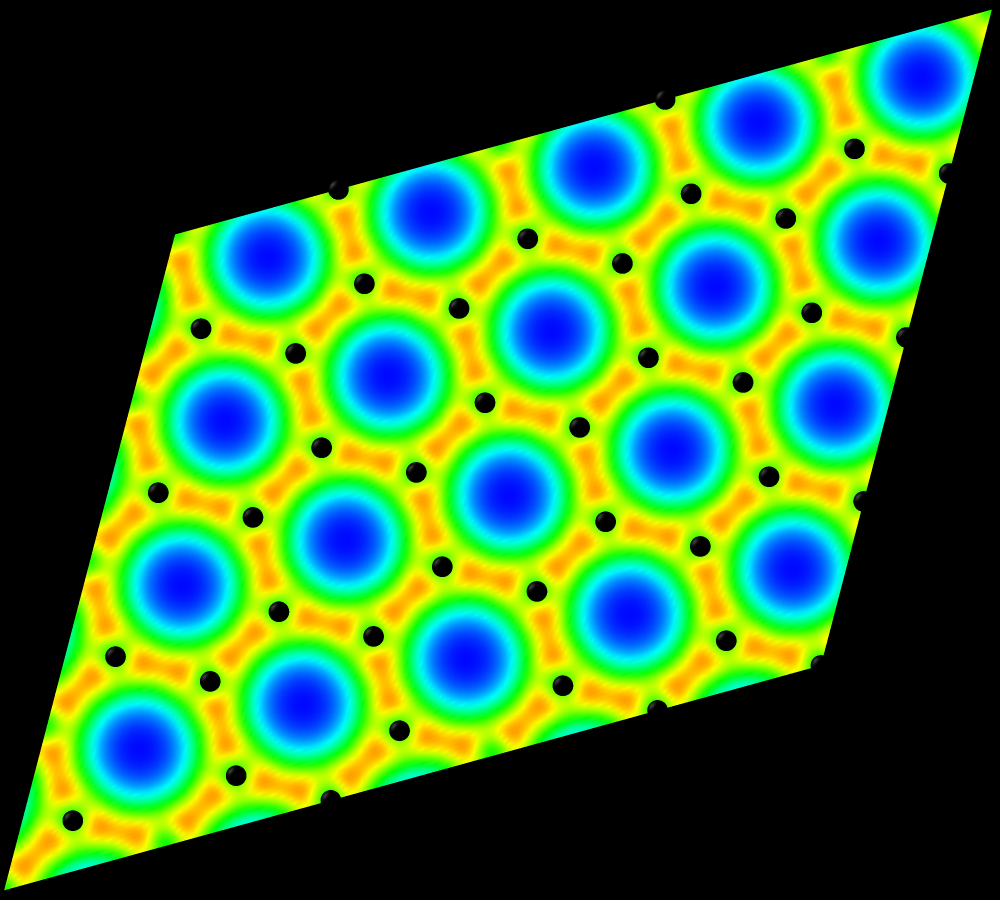}
		\end{subfigure}

		\caption{{\bf (top)} Charge density distribution on two neighboring graphitized planes for M2 model simulated at 2700K and three equally spaced 
		slices between them. {\bf (bottom)} Similar illustration for cG (lower panel) is included for the purpose of comparison. Black circles in (a) and (e) 
		mark the position of the atoms in the plane. }
		\label{fig:slab_CHG}
	\end{adjustbox}
\end{figure*}

\begin{figure*} [ht]
	\begin{adjustbox}{varwidth = \textwidth, center,   rndfbox= 3ex 3ex 3ex 3ex}
      		\begin{minipage}[b]{.95\textwidth}
			\centering
		 	\includegraphics[width=0.990\textwidth]{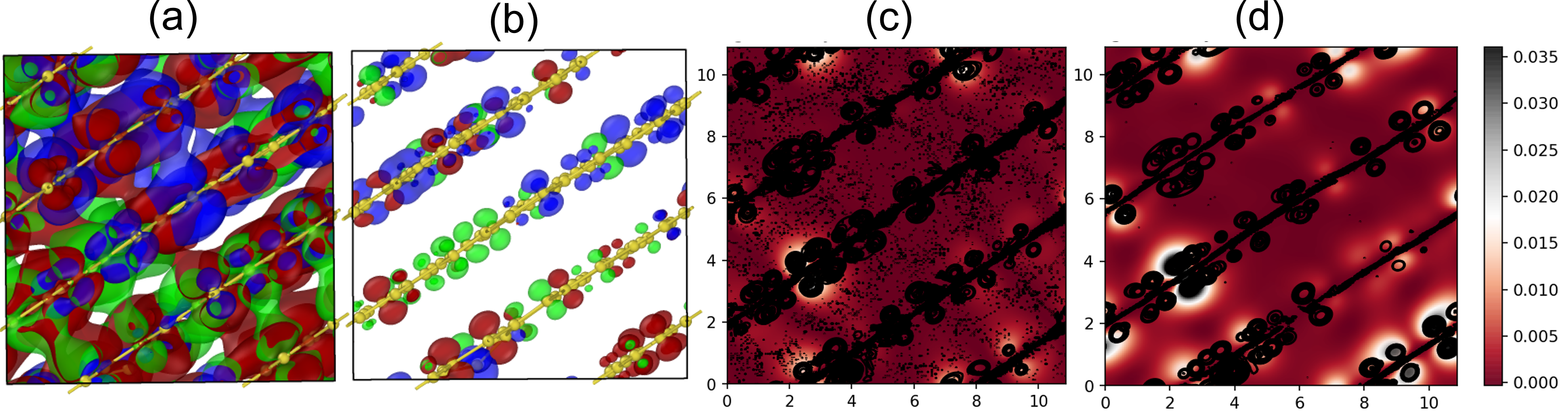} 
		\end{minipage}
   		\caption{Details of band-decomposed charge densities for the a-G for (a) 3 $\pi$-bands (colored blue, green, and red) in the valence region and (b) their correspondingly symmetric $\pi$*-bands 
		in the conduction region. We also show the charge distribution for a pair of nearly-symmetric bands in (c) the valence region and (d) the conduction region.}
   		\label{fig:ChargeDensity} 
	\end{adjustbox}
\end{figure*} 

\begin{figure*} [ht]
	\begin{adjustbox}{varwidth = \textwidth, center,   rndfbox= 3ex 3ex 3ex 3ex}
      		\begin{minipage}[b]{.95\textwidth}
			\centering
		 	\includegraphics[width=0.990\textwidth]{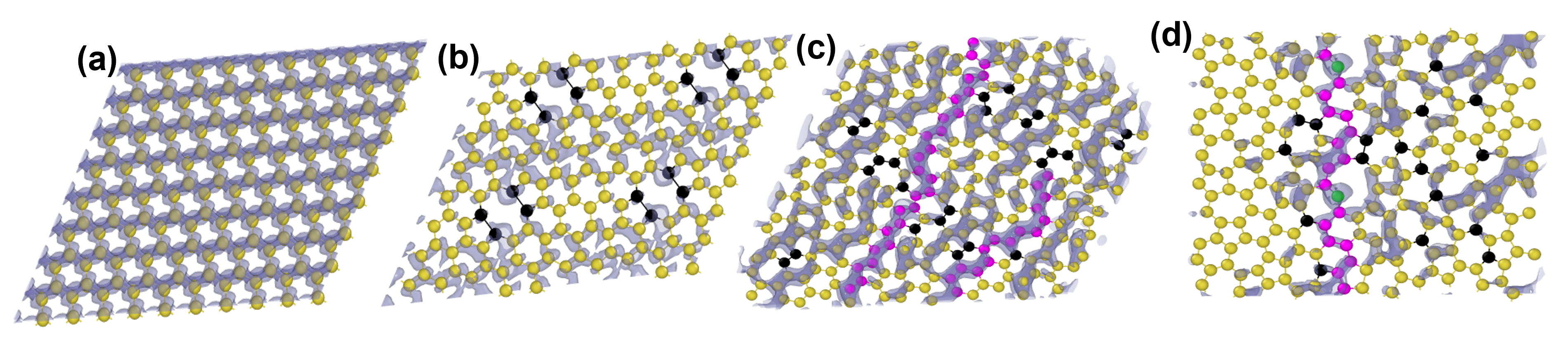} 
		\end{minipage}
   		\caption{SPC results (grey isosurface) for (a) an ideal graphene layer, (b) a graphene layer with 2 vacancies, (c) M1, and (d) M2. Atoms in pink in (c) and (d) show atoms forming conduction paths in the spatial grid, while atoms in black are border atoms where one or both neighboring rings are non-hexagon rings. Green colored atoms in (d) are consistent for SP atoms in M2.}
   		\label{fig:SPC} 
	\end{adjustbox}
\end{figure*} 

The charge-density distribution for the M2 model simulated at 2700K has been presented, together with similar calculations for c-G for comparison, in Fig. \ref{fig:slab_CHG}. The charge distribution was calculated using the Heyd, Scuseria and Ernzerhof (HSE06) hybrid functional \cite{HSE06, HSE06_I,HSE06_II} and has been plotted along two neighboring planes of atoms (labelled `a', `e') and three other parallel, equally spaced slices (labelled `b', `c', `d') in between them (in the gallery). 
For comparison, respective planes in a-G and cG have been plotted within the same color range. For plane (`c'), the colormap shows contributions from both planes. 
The color maps for a-G show a more disordered distribution of charges along the planes of atoms, compared to the cG, particularly because of the presence of bond-length/bond-angle distortion, ring disorder induced puckering, etc. Our calculations have also indicated that the variation of the charge density values for the a-G is higher than graphite because of the disorder. The charge distribution in the a-G galleries exhibits a low-density delocalized electron gas with higher charge on the plane of atoms and monotonically decreasing as we move away into the gallery. However, we should note that the majority of the charge density on the most isolated layer from the plane of atoms, plane `c' in Fig. \ref{fig:slab_CHG}, is greater than 2\% of the maximum charge density on the plane of atoms (layer `a' and `e' ), suggesting the presence of a fairly homogeneous electron gas in the galleries built from the bonding orbitals formed from the $\pi$-electrons. The electronic density of states (DoS) of a-G revealed a broad peak at the Fermi level and had no semi-metallic DoS characteristic of c-G.

In Fig. \ref{fig:ChargeDensity}, we present information on the band-decomposed charge densities for the a-G. Bands close to the Fermi level ($E_f$) contain the $\pi$ ($E < E_f$) and $\pi$* ($E > E_f$) electrons. The $\pi$-bands involve much mixing from $\pi$-orbitals on different sites. Fig. \ref{fig:ChargeDensity}a shows the $\pi$ mixing for 3 $\pi$-bonding orbitals, The $\pi$-electrons from these bands extend into the gallery, creating binding between layers separated by roughly 3.1 $\mathring{A}$.  Fig. \ref{fig:ChargeDensity}b shows the $\pi$* anti-bonding orbitals with no charge projection to the gallery. The evidence of the electron delocalization is further illustrated by projecting the charge density from the 3d box into a plane for a single $\pi$ band and a  symmetric $\pi$* band in Fig. \ref{fig:ChargeDensity}c and d, supporting the presence and absence of the charge density in the gallery for the $\pi$ and $\pi$* orbitals respectively.  The presence of such delocalized $\pi$-electrons in the galleries has been suggested for graphite; where it was argued that the graphene bonding forces are  dominantly metallic and not Van der Waals \cite{Rozploch07, Santos72}. Our work suggests that while Van der Waals plays a role in layering and binding, other contributions within LDA or PBE {\it also play an important role}.
 
To study the effects of disorder on the electronic conduction and visualize the conduction-active regions in the network, we calculate the space-projected conductivity (SPC) \cite{KashiPSSb} on a-G and compare it with that of c-G. The SPC exploits the Kubo-Greenwood formula to obtain information about conduction pathways in materials. The SPC projected onto particular layers of atoms are shown in Fig.\ref{fig:SPC}. The SPC of an ideal graphite layer with no defects has clear paths for conduction in the plane. However, in graphite with a 5-8-5 ring defect, the conduction in the regions connecting the pentagons with the octagon is significantly reduced but the conductivity is still high in the regions dominated largely by hexagons. This reduction in electric conductivity in a 5-8-5 defected graphene has been previously reported \cite{Liu19}. Similar findings were seen for our atomic layers in a-G whereby the conduction paths try to avoid a junction involving a ring disorder. In other words, conduction is favored along connected atoms in hexagonal rings over non-hexagon rings. The presence of topological ring disorder significantly affects the charge transport in both graphite and a-G. We also found that the conductivity value in pure graphite is highest, followed by 5-8-5 defect graphite. The a-G conductivity was decreased by a factor of about $10^{-2}$ relative to graphite.  

In conclusion, we present evidence that a-G exists and we describe its process of formation. Plane formation is found to be robust in a suitable temperature/density window.  a-G growth may be a practical means to obtain amorphous graphene planes in a layered graphite-like superstructure, that might even be exfoliated. We expect the in-plane electronic conductivity to be much reduced compared to graphite and expect this to be another signature of a-G. We analyze the electronic structure, the mechanism of cohesion and compute the electronic consequences of topological (ring) disorder using the space-projected conductivity.

We thank S. R. Elliott, H. Castillo, B. Bhattarai, E. A. Stinaff, M. E. Kordesch, M. F. Thorpe,  and the US Dept. of Energy for support under grant DE-FE0031981 and XSEDE (supported by National Science Foundation grant number ACI-1548562) for computational support under allocation number DMR-190008P. 

\newpage
\bibliography{reference}
\bibliographystyle{unsrt}
\end{document}